\documentclass[%
 preprint,
 superscriptaddress,
 longbibliography,
 amsmath,amssymb,
 aps, physrev,
]{revtex4-2}

\usepackage{graphicx}
\usepackage{dcolumn}
\usepackage{bm}
\usepackage{epstopdf}
\usepackage{adjustbox}
\usepackage[utf8]{inputenc}
\usepackage[T1]{fontenc}
\usepackage{mathptmx}
\usepackage{upgreek}
\usepackage{textcomp}
\usepackage{gensymb}
\usepackage{anyfontsize}
\usepackage{amsmath}
\usepackage{hyperref}
\usepackage{natbib}
\usepackage{siunitx}
\usepackage{physics}
\usepackage{float}

\newcommand{\crcl}{CrCl$_3$}
\newcommand{\FDT}{$\frac{d}{dB} |S_{21}|^{2}$}
\newcommand{\FDR}{$\frac{d}{dB} |S_{11}|^{2}$}

\begin{document}

\title{Cavity based sensing of antiferromagnetic canting and nonzero-momentum spin waves in a van der Waals cavity-magnon-polariton system}

\author{Supriya Mandal}
\email{supriyam@illinois.edu}
\thanks{Corresponding Author}
\affiliation{University of Illinois Urbana-Champaign, Urbana, Illinois 61801, USA}
\affiliation{Tata Institute of Fundamental Research, Mumbai 400005, Maharashtra, India}
\author{Krishnendu Maji}
\affiliation{Tata Institute of Fundamental Research, Mumbai 400005, Maharashtra, India}
\author{Lucky N. Kapoor}
\affiliation{Institute of Science and Technology Austria, Am Campus 1, 3400 Klosterneuburg, Austria}
\affiliation{Tata Institute of Fundamental Research, Mumbai 400005, Maharashtra, India}
\author{Souvik Sasmal}
\affiliation{Tata Institute of Fundamental Research, Mumbai 400005, Maharashtra, India}
\author{Soham Manni}
\affiliation{Indian Institute of Technology Palakkad, Kanjikode, Palakkad 678623, Kerala, India}
\author{John Jesudasan}
\affiliation{Tata Institute of Fundamental Research, Mumbai 400005, Maharashtra, India}
\author{Pratap Raychaudhuri}
\affiliation{Tata Institute of Fundamental Research, Mumbai 400005, Maharashtra, India}
\author{Arumugam Thamizhavel}
\affiliation{Tata Institute of Fundamental Research, Mumbai 400005, Maharashtra, India}
\author{Mandar M. Deshmukh}
\email{deshmukh@tifr.res.in}
\thanks{Corresponding Author}
\affiliation{Tata Institute of Fundamental Research, Mumbai 400005, Maharashtra, India}

\date{\today}

\begin{abstract}

    Cavity-magnon-polaritons are hybrid excitations from the interaction between cavity photons and magnons, the quanta of collective spin oscillations. Along with the tunability of the magnon-photon coupling strength, fast information transfer and conversion speed are desired in hybrid devices. This can be achieved utilizing the propagating nature of spin waves with non-zero momentum for their ultra-fast time dynamics and reduced ohmic dissipation. Antiferromagnets are particularly interesting as hosts for magnons since stray-field interactions are minimized, and they support multiple modes with distinctive magnetic-field behavior across the phase diagram. Chromium trichloride (CrCl$_{3}$) is a van der Waals layered antiferromagnet having a strong easy-plane anisotropy and a weak in-plane easy-axis anisotropy. Despite some magnetic resonance studies, the impact of magnetic reorientation of spins in CrCl$_3$ on the cavity-magnon-polariton interaction strength as a function of magnetic field remains largely unexplored. In this study, we investigate the coupling between magnons in CrCl$_{3}$ and photons in a coplanar waveguide resonator as a function of magnetic field. In particular, we find that the magnon-photon coupling strength varies nonmonotonically and distinctly with the magnetic field for the acoustic and the optical magnons, which can be utilized to tune the magnon-photon coupling strength using an external magnetic field as a knob. We find the signature of spin-flop transition in the two harmonics of the cavity due to a stronger dispersive coupling between optical magnons and cavity photons at lower fields. Additionally, we find standing modes formed by spin waves with nonzero momentum associated with the two hybrid magnons when the external field is applied at an angle with the crystal plane. These modes do not undergo substantial coupling with the cavity mode unlike the antiferromagnetic modes and can be used as low-loss propagation channels in hybrid devices.
\end{abstract}

\maketitle

\renewcommand{\thesection}{\Roman{section}}
\setcounter{section}{0}

\renewcommand{\thefigure}{\arabic{figure}}
\setcounter{figure}{0}

\renewcommand{\theequation}{\arabic{equation}}
\setcounter{equation}{0}

\renewcommand{\thetable}{\Roman{table}}
\setcounter{table}{0}

\section{Introduction}

Cavity-magnonics studies the hybrid modes formed when an electromagnetic mode in a cavity interacts with a collective spin oscillation mode in a magnetic material. The quantum of such a hybrid mode is called a cavity-magnon-polariton. Traditional application of magnonics typically centers around a ferromagnet (FM) or a ferrimagnet with a low damping constant, the most prominent being yttrium iron garnet (YIG). Over the years, several useful devices, such as microwave circulators, filters and other magnonic devices, have been made possible by this \cite{flebusRecentAdvancesMagnonics2023,branfordMaterialsPhysicsDevices2024,hanMagnonicsMaterialsPhysics2024}. In recent times, increasing complexity in magnon-based hybrid quantum architecture demands additional knobs concerning magnetic field symmetry and tunability of coupling strength between photons and magnons. Antiferromagnets (AF) hold the potential to meet this requirement because of their complex magnetic phase diagram, competing anisotropies, and multiple resonance modes with distinctive dynamics. This potential is further enhanced with the emergence of van der Waals type magnetic materials including antiferromagnets with unique properties \cite{khanSpinDynamicsStudy2019, cenkerDirectObservationTwodimensional2021, mcguireMagneticBehaviorSpinlattice2017, zhangRobustIntrinsicFerromagnetism2015, zhangGatetunableSpinWaves2020, caiAtomicallyThinCrCl32019, macneillGigahertzFrequencyAntiferromagnetic2019}. To integrate antiferromagnets into hybrid device architectures, understanding how magnetic field affects their magnetic structure and magnon-photon coupling strength becomes essential.

\begin{figure}[H]
    \centering
    \includegraphics{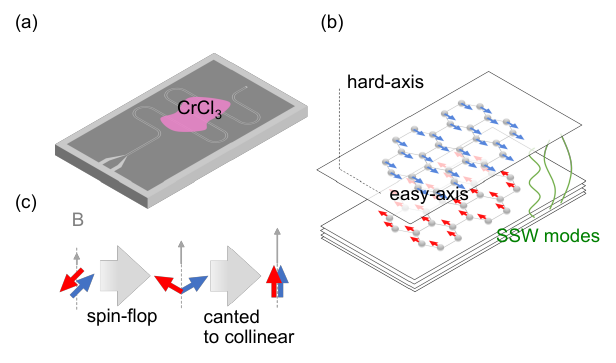}
    \caption{(a) Schematic showing CrCl$_3$ crystal placed on a single-port NbN CPW resonator for measuring magnon-photon coupling in reflected signal. (b) Schematic showing van der Waals stacking of atomic layers in CrCl$_3$ along with out-of-plane hard axis and in-plane easy axis. The spacing between the first two layer have been exaggerated to show the oppositely oriented spins in adjacent layers due to inter-planar antiferromagnetic exchange. Standing spin wave (SSW) modes formed across the thickness of the crystal are schematically shown in green lines. (c) Schematic showing the spin-flop transition governing the transformation of adjacent sublattice moments from antiferromagnetic to canted orientation, and the transition at higher field from canted to a collinear ferromagnet-like alignment.}
    \label{fig:fig1}
\end{figure}

In an antiferromagnet, direction-dependent magnetic anisotropy can occur due to structural effects like dipolar interactions, crystal-field, and spin-orbit interactions \cite{machadoSpinflopTransitionEasyplane2017}. The presence of competing hard and soft anisotropy axes can result in a two-sublattice (TSL) AF, such as in the case of NiO \cite{machadoSpinflopTransitionEasyplane2017, rezendeIntroductionAntiferromagneticMagnons2019, liensbergerExchangeEnhancedUltrastrongMagnonMagnon2019, macneillGigahertzFrequencyAntiferromagnetic2019}. In the case of chromium trichloride (\crcl), a layered antiferromagnet that has garnered recent interest for showing rich antiferromagnetic mode dispersions at gigahertz frequencies \cite{macneillGigahertzFrequencyAntiferromagnetic2019, kapoorObservationStandingSpin2021, mandalCoplanarCavityStrong2020, zhangCouplingMicrowavePhotons2021, zhangZerofieldMagnonPhoton2021, chenMasslessDiracMagnons2021, lyonsAcousticallyDrivenMagnonPhonon2023, sunTemperaturecontrolledStrongMagnon2025}, the dominant determining factor for the magnetic anisotropy is its cleavable van der Waals structure. During the cooling stage in the preparation of a \crcl{} single crystal, first, an intra-planar FM structure appears with exchange energy of 0.5~meV; afterward, these FM atomic planes get oriented in AF fashion in the third dimension resulting in a weaker inter-planar AF structure with exchange energy of 1.6~\micro eV \cite{kuhlowMagneticOrderingCrCl31982, mcguireMagneticBehaviorSpinlattice2017, macneillGigahertzFrequencyAntiferromagnetic2019}. This makes \crcl{} a two-sublattice AF with a low overall anisotropy such that a few kOe field is sufficient enough to rotate the magnetization in almost any direction. The structure of \crcl{} is shown schematically in Fig~\ref{fig:fig1}~(b).  The low anisotropy, in-plane FM order, and cleavable van der Waals structure make \crcl{} particularly interesting for coupling with other materials in a hybrid device scheme.

When a magnetic field is applied parallel to the crystal plane, a TSL AF with low magnetic anisotropy undergoes two main transitions, as shown in Fig~\ref{fig:fig1}~(c). At very low fields, the individual moments stay in an antiparallel orientation. As the field is increased, the two sublattice moments undergo a sudden transition to a canted state where the moments make a non-zero angle (canting angle $\theta$) with the applied field direction when the field is parallel to the initial direction of the moments; this is called spin-flop transition in an AF. However, if the field is perpendicular to the moments, a gradual transition to the canted state occurs. As the field increases more, the canting angle decreases. Beyond another critical field, all the moments get oriented in a collinear FM-like fashion. The spin-flop field marks two distinct regimes when the system is in AF state. As a result, the nature of the collective spin excitations in the system also changes across the spin-flop field. The spin-flop transition often takes place at a very low field where the effective dipole moment is weak in an AF. Although spin-flop transition has been studied for the static magnetic structure of different AFs, its effect on collective spin dynamics in such systems remains largely unexplored. In this study, we show that the spin-flop and additional spin dynamics in \crcl{} can be studied by measuring harmonics of a cavity coupled to the antiferromagnetic resonance (AFMR) modes in \crcl. The sensitive nature of a cavity-based measurement in resonant and dispersive regimes of magnon-photon interaction provides insights into the evolving nature of the magnetic structure, spin dynamics, and magnon-photon coupling strength with field.

In addition to the AFMR modes, standing spin wave (SSW) modes can form due to the presence of boundary conditions across the thickness of the few micron thick CrCl$_{3}$ crystal, causing propagating spin waves to undergo reflections, as shown schematically in Fig~\ref{fig:fig1}~(b). Contrary to the long wavelength AFMR modes, these spin waves have nonzero momentum and are propagating in nature. These spin waves with $k \ne 0$ ($k$ is the magnitude of the wave vector) form when adjacent spins start precessing with a finite phase lag between each other. These modes appear as a multiplicity of modes in the vicinity of the AFMR modes \cite{kittelExcitationSpinWaves1958,kapoorObservationStandingSpin2021, seaveyDirectObservationSpinWave1958, luiAntiferromagneticStandingspinwaveResonance1990}. We observe these higher order SSW modes corresponding to both the hybrid AFMR modes resulting from hybridization between the acoustic and the optical modes for a finite field angle with respect to the crystal plane. After the transition to the collinear state, we also see them associated with the FMR-like mode. Interestingly, we find that although the AFMR modes couple strongly to the cavity mode, the SSW modes stay largely decoupled. This shows that the SSW modes can be useful as low-dissipation information transfer channels in situations where non-interacting propagating modes are desired in a hybrid otherwise coupled device scheme.

\section{Field variation of magnon-photon coupling strength}

Our device consists of a 20-30~\unit{\micro\metre} thick CrCl$_3$ crystal placed at the current anti-node of a half-wave coplanar waveguide (CPW) resonator made of 80~nm thick niobium nitride (NbN) deposited on intrinsic silicon substrate, as shown in a schematic in Fig~\ref{fig:fig1}~(a). The type-II disordered superconductor, NbN, is used to make the resonators field compatible \cite{mandalCoplanarCavityStrong2020}. This helps in minimizing the losses as well as reducing cavity frequency shift when field is applied parallel to the device plane. We use an optical microscope based micromanipulation technique to place the crystal on top of the NbN resonator in a nitrogen environment inside a glove-box and coat with a layer of Apeizon N grease to avoid quality degrading ambient effects on the crystal. We mount the device on a rotatable sample holder to load in an Oxford Triton dilution fridge. All the measurements have been done at a temperature of 20~mK. For applying magnetic field, we have used a superconducting coil scalar magnet inside the fridge. We have used Anritsu MS46122B vector network analyzer (VNA) to measure the reflection coefficient, $S_{11}$, of the signal reflected from our single-port device as a function of frequency. For the majority of our study, we have used the magnetic field derivative of the measured $S_{11}$ which brings out the features in the frequency domain that vary with the magnetic field, suppressing noise and frequency dependent background variation.

\begin{figure}[H]
    \centering
    \includegraphics{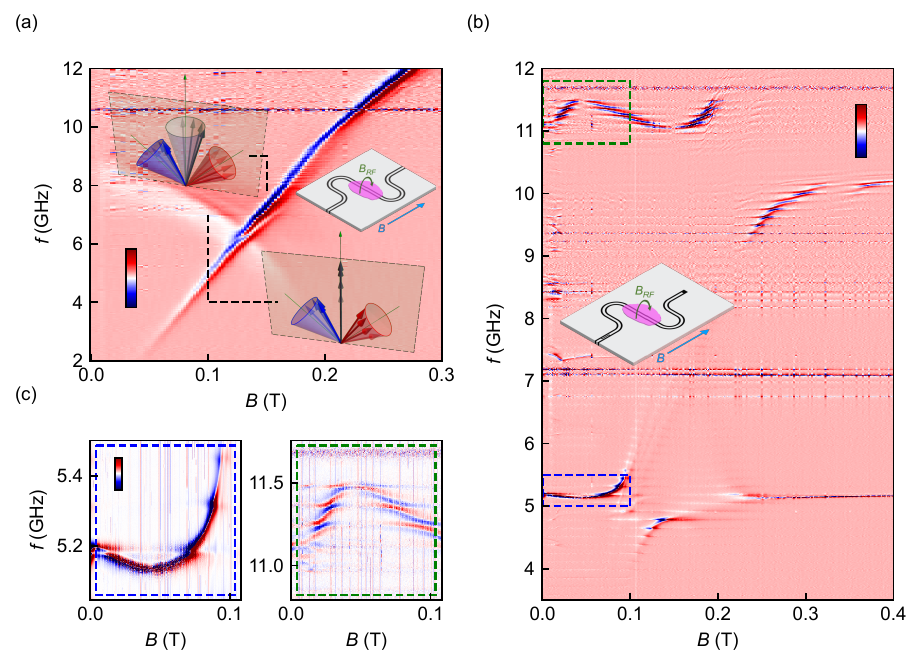}
    \caption{(a) Field derivative of transmission spectra of \crcl{} on a transmission line as a function of frequency and in-plane magnetic field applied in an orientation shown in the inset (color-bar shows \FDT{} ranging from -40 to +30~dB/T). The acoustic (linear) and the optical (quadratic) AFMR modes are visible. The insets show a timelapse sketch of the orientation and the resultant moment of the two sublattice moments for the corresponding modes (indicated by dashed lines). (b) Field derivative of reflection spectra of \crcl{} on a CPW resonator as a function of frequency and magnetic field (colorbar shows \FDR{} ranging from -40 to +30~dB/T) for a similar field orientation, shown in the inset. Two avoided crossings for coupling of the lower harmonic of the cavity with the acoustic and optical AFMR modes are visible around 5~GHz, and one avoided crossing for coupling of the upper harmonic of the cavity with the FMR mode is visible around 10~GHz following the transition of \crcl{} into collinear FM-like state. (c) The regions marked by a blue and a green dashed rectangle in (b) have been enlarged to show the opposite-ward bend of the lower and the upper cavity mode around 40~mT (colorbar ranges from -100 to +100~dB/T here).}
    \label{fig:fig2}
\end{figure}

Measurement of the reflection coefficient of the one-port cavity coupled to the antiferromagnet as a function of frequency and magnetic field shows four distinct modes: fundamental and first harmonic of the cavity mode, and two modes associated with CrCl$_3$: acoustic and optical AFMR modes. We have shown these two AFMR modes from a separate transmission spectroscopy done on a CrCl$_{3}$ crystal placed on a CPW transmission line in Fig.~\ref{fig:fig2}~(a) along with the sublattice precession corresponding to the modes in the inset. For the hybrid cavity-antiferromagnet system, when the magnetic field is increased, at a first glance we observe two prominent features (shown in Fig.~\ref{fig:fig2}~(b)); namely, two avoided crossings as the acoustic and the optical AFMR modes separately hybridize due to resonant interaction with the fundamental mode of the cavity, and a single larger avoided crossing due to hybridization of the FMR mode after the transition to the collinear state with the first harmonic of the cavity. 

This is expected and shows that there exists magnon-photon coupling between the cavity photons and the magnons corresponding to each of the AFMR modes. As the frequencies of each of the cavity and AFMR mode pairs become equal as a result of mode dispersion with magnetic field (here we neglect the mild dispersion of the cavity as the field is applied in plane \cite{mandalCoplanarCavityStrong2020}), two avoided crossings show up. They are quantified by magnon-photon coupling strengths between the cavity mode and each of the acoustic and the optical AFMR modes, denoted by $g_\textrm{ac}$ and $g_\textrm{op}$ respectively. On closer inspection, we notice another subtle feature: around 15-20 mT field, the fundamental cavity mode starts bending downward by an amount $\sim$ 70~MHz whereas the first harmonic bends upward by an amount $\sim$ 400~MHz. This is shown in the zoomed plots in Fig.~\ref{fig:fig2}~(c). It is curious to note that this field is quite close to the spin-flop transition field in CrCl$_3$ \cite{kuhlowMagneticOrderingCrCl31982, mcguireMagneticBehaviorSpinlattice2017, caiAtomicallyThinCrCl32019}. We note that deviation from standard dispersion close to spin-flop field has been reported in another literature recently \cite{evertsUltrastrongCouplingMicrowave2020}. Across spin-flop transition a TSL AF undergoes a transition from an anti-parallel AF state to a canted spin orientation. This change in magnetic microstructure affects the response of the device. Firstly, this is accompanied by a change in the mode dispersion as a function of field. But we find that this effect is small enough to not cause more than a fraction of a MHz change in the cavity frequency (shown in the Supplemental Material \cite{SupplMat}). Another parameter that should get largely affected by a change in the magnetic structure of a material is the magnon-photon coupling strength itself. Substantial opposite-ward bending of the two cavity harmonic modes around the spin-flop transition suggests a dispersive coupling between the optical AFMR mode and the cavity modes. To understand this, we write down the Hamiltonian of the system with cavity-spin Zeeman coupling, exchange coupling between neighboring spins in the antiferromagnet, and the anisotropies corresponding to both the hard (perpendicular to the crystal plane) and the easy axis (in the crystal plane)

\begin{equation}
    H = -\gamma \hbar \sum_{i,j} \vb{H}_0 \vdot \vb{S}_{i,j} + \sum_{i,j} 2 J_{ij} \vb{S}_{i} \vdot \vb{S}_{j} +\sum_{i,j} D_{y} (S_{i,j}^{y})^{2} - \sum_{i,j} D_{z} (S_{i,j}^{z'})^{2}
\end{equation}

where $\vb{H}_0$ is the external field, $\vb{S}_{i(j)}$ is the spin at the $i$th site of 1st ($j$th site of 2nd) sublattice, $J_{ij}>0$ is the antiferromagnetic exchange constant, and $D_{y}$ and $D_{z}$ are anisotropy strengths corresponding to the hard and the easy axis. Here we take the hard axis to be the $y$-axis which is perpendicular to the crystal plane and $z'$ to be an easy-axis on the crystal plane. Next, following Holstein-Primakoff transformation \cite{holsteinFieldDependenceIntrinsic1940, boventerAntiferromagneticCavityMagnon2023, rezendeIntroductionAntiferromagneticMagnons2019, machadoSpinflopTransitionEasyplane2017}, we express the spin operators in terms of spin wave excitation or magnon operators $S_{i}^{x'} = \sqrt{\frac{S}{2}} (a_{i} + a_{i}^{\dag})$, $S_{i}^{y'} = \frac{1}{i} \sqrt{\frac{S}{2}} (a_{i} - a_{i}^{\dag})$ and $S_{i}^{z'} = S - a_{i}^{\dag} a_{i}$, which follow the commutation relations $\left[ a_{i}, a_{i'}^\dagger \right] = \delta_{i,i'}$, $\left[ a_{i}, a_{i'} \right] = 0$, $\left[ b_{j}, b_{j'}^\dagger \right] = \delta_{j,j'}$, and $\left[ b_{j}, b_{j'} \right] = 0$ (for spins in the other sublattice, we replace $a$ by $b$, and $x'$, $y'$, $z'$ by $x''$, $y''$, $z''$ respectively). In the canted phase, applying a rotation operation (by angles $\pm \theta$ for either sublattice) on the sublattice spin operators, and expressing the magnon creation and annihilation operators in their Fourier components as $a_{i} = \frac{1}{\sqrt{N}} \sum_{k} {e^{i \vb{k}.\vb{r}_{i}} a_{k}}$ and $b_{j} = \frac{1}{\sqrt{N}} \sum_{k} {e^{i \vb{k}.\vb{r}_{j}} b_{k}}$ where $N$ represents number of spins in each sublattice, and the operators follow the commutation relations $\left[ a_{k}, a_{k'}^\dagger \right] = \delta_{k,k'}$, $\left[ a_{k}, a_{k'} \right] = 0$, $\left[ b_{k}, b_{k'}^\dagger \right] = \delta_{k,k'}$, and $\left[ b_{k}, b_{k'} \right] = 0$, we obtain the following form of the Hamiltonian

\begin{equation}
\begin{split}
    H &= \sum_{\eta} H_{\eta} \\
    H_{1} &= \gamma \hbar H_{0} \sum_{k} (a_{k}^{\dag} a_{k} + b_{k}^{\dag} b_{k}) \cos{\theta} \\
    H_{2} &= 
    \begin{aligned}[t]
        &\gamma \hbar H_{E} \sum_{k} [(a_{k}^{\dag} a_{k} + b_{k}^{\dag} b_{k}) \cos{2\theta} \\ & + (a_{k} b_{k}^{\dag} + a_{k}^{\dag} b_{k}) \cos^{2}{\theta} \\ & - (a_{k}^{\dagger} b_{-k}^{\dag} + a_{k} b_{-k}) \sin^{2}{\theta}]
    \end{aligned} \\
    H_{3} &= 
    \begin{aligned}[t]
        -&\gamma \hbar \frac{H_{Ay}}{4} \sum_{k} [(a_{k} a_{-k} + b_{k} b_{-k} + h.c.) \\ - & 2 (a_{k}^{\dag} a_{k} + b_{k}^{\dag} b_{k})]
    \end{aligned} \\
    H_{4} &= 
    \begin{aligned}[t]
        &\gamma \hbar \frac{H_{Az}}{4} \sum_{k} [(a_{k} a_{-k} + b_{k} b_{-k} + h.c.) \sin^{2}{\theta} \\ & + 2 (\sin^{2}{\theta} - 2 \cos^{2}{\theta}) (a_{k}^{\dag} a_{k} + b_{k}^{\dag} b_{k})]
    \end{aligned}
\end{split}
\end{equation}

\begin{figure}[H]
    \centering
    \includegraphics{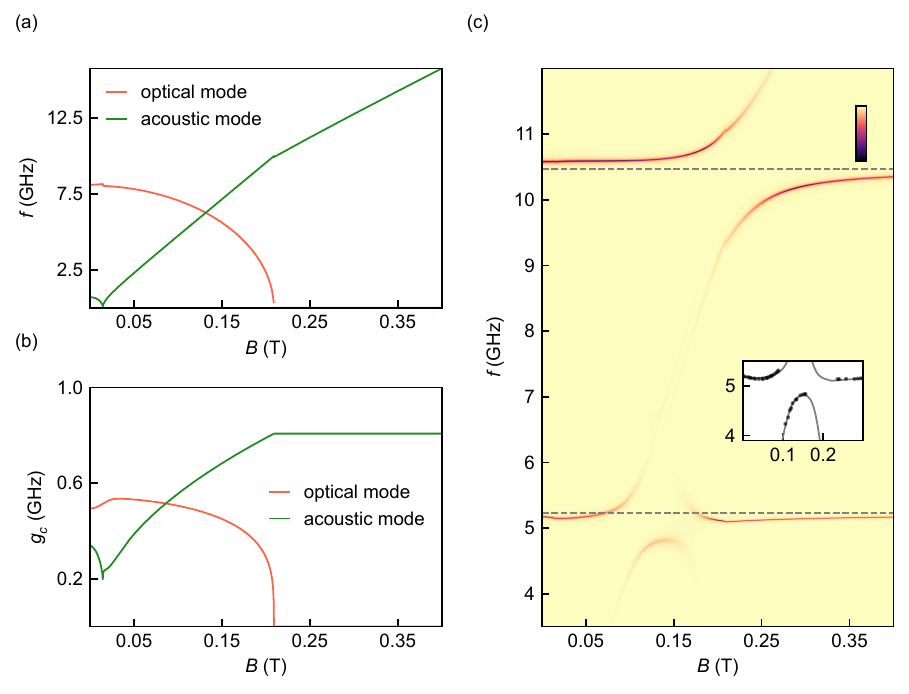}
    \caption{(a) Magnetic field dependence of frequencies of the optical and the acoustic AFMR modes as calculated using the TSL AF model. (b) Magnetic field dependence of the magnon-photon coupling strengths of the optical and the acoustic AFMR modes with the cavity mode calculated using the TSL AF model and using fit parameters from the fit to the mode dispersion. (c) Reflection coefficient of the full cavity antiferromagnet system calculated from the input-output theory using the field dependencies of the frequency and coupling strength as shown in (a) and (b) (colorbar ranges from -25 to 0 dB) (inset: fit to the mode dispersion using the TSL AF model described in the text).}
    \label{fig:fig3}
\end{figure}

Here $H_{E}$ is the exchange field, $H_{Ay}$ is the hard-axis anisotropy and $H_{Az}$ is the easy-axis anisotropy. They can be expressed in terms of the spin parameters as $H_{E} = \frac{2 J S}{\gamma \hbar}$, $H_{Ay} = \frac{2 D_{y} S}{\gamma \hbar}$, and $H_{Az} = \frac{2 D_{z} S}{\gamma \hbar}$. The angle $\theta$ is the canting angle that the sublattice moments make with the direction of the magnetic field and has a field dependence $\theta (H) = \cos^{-1}{\left( \frac{H}{2 H_{E} - H_{Az}} \right)}$. Since we mainly focus on the lowest order modes, we can set $k=0$. Then the above equation simplifies to
\begin{equation}
    H = A (a^{\dag} a + b^{\dag} b) + B (a b + a^{\dag} b^{\dag}) + \frac{C}{2} (a^{2} + b^{2} + \textrm{h.c.}) + D (a b^{\dag} + a^{\dag} b)
\end{equation}
where
\begin{equation}
    \begin{split}
        A &= \gamma \left[ H_{0} \cos{\theta} - H_{E} \cos{2 \theta} + \frac{1}{2} H_{Ay} + \frac{1}{2} H_{Az} \left( 2 \cos^{2}{\theta} - \sin^{2}{\theta} \right) \right] \\
        B &= -\gamma H_{E} \sin^{2}{\theta} \\
        C &= -\gamma \left( \frac{H_{Ay}}{2} + \frac{H_{Az}}{2} \sin^{2}{\theta} \right) \\
        D &= \gamma H_{E} \cos^{2}{\theta}
    \end{split}
    \label{eq:ABCD}
\end{equation}
and h.c. represents hermitian conjugate. Next, we diagonalize this Hamiltonian to find the normal modes which we denote by $\alpha$ and $\beta$. They are related to $a$ and $b$ through
\begin{equation}
    \mqty(a \\ b \\ a^{\dag} \\ b^{\dag}) = \mqty(m & n & p & q \\ -m & n & -p &q \\ p & q & m & n \\ -p & q & -m & n) \mqty(\alpha \\ \beta \\ \alpha^{\dag} \\ \beta^{\dag})
    \label{eq:abmatrix}
\end{equation}
where
\begin{equation}
    \begin{split}
        m &= \sqrt{\frac{(A - D) + \omega_{\alpha}}{4 \omega_{\alpha}}} \\
        n &= \sqrt{\frac{(A + D) + \omega_{\beta}}{4 \omega_{\beta}}} \\
        p &= \sqrt{\frac{(A - D) - \omega_{\alpha}}{4 \omega_{\alpha}}} \\
        q &= \sqrt{\frac{(A + D) - \omega_{\beta}}{4 \omega_{\beta}}} \\
    \end{split}
    \label{eq:mnpq}
\end{equation}
with
\begin{equation}
    \begin{split}
        \omega_{\alpha}^{2} &= (A - D)^{2} - (B - C)^{2} \\
        \omega_{\beta}^{2} &= (A + D)^{2} - (B + C)^{2}
    \end{split}
\end{equation}
These $\omega_{\alpha}$ and $\omega_{\beta}$ are the eigenfrequencies of the system corresponding to the optical and the acoustic AFMR modes. Their field dependencies are calculated and shown in Fig.~\ref{fig:fig3}~(a) using the parameters for CrCl$_{3}$ $H_{E} = 105$~mT, $H_{Ay} = 396$~mT, $H_{Az} = 1$~mT and $\gamma / 2\pi = 28$~GHz/T \cite{macneillGigahertzFrequencyAntiferromagnetic2019, mcguireMagneticBehaviorSpinlattice2017}.
Using equations Eq.~\ref{eq:ABCD} and Eq.~\ref{eq:mnpq}, along with the standard expression for the magnetic moment vector $\vb{m} = \sum_{\eta} \vu{\eta} m_{\eta}$ where $m_{\eta} = \sqrt{N} \left( S_{i}^{\eta} + S_{j}^{\eta} \right)$ and $\eta = x, y, z$, we obtain the following expressions for the magnitudes of the magnetic moments
\begin{equation}
    \begin{split}
        m_{x} &= 2 (n + q) (\beta + \beta^{\dag}) K \cos{\theta} \\
        m_{y} &= 2 (n - q) (\beta - \beta^{\dag}) K \\
        m_{z} &= 2 (m + p) (\alpha + \alpha^{\dag}) K \sin{\theta}
    \end{split}
\end{equation}
where $K=\sqrt{NS/2}$. Here $\alpha$ and $\beta$ represent the optical and the acoustic modes respectively. From these expressions and from the geometry of the coplanar waveguide, shown in the inset of Fig.~\ref{fig:fig2}~(b), we find that on the central line the $z$-component of the moment couples to the corresponding polarization vector component of the microwave photons giving rise to the coupling between microwave photons and optical magnons. Whereas, in the gaps on either side the $y$-component of the moment couples to the photon polarization giving rise to coupling between microwave photons and acoustic magnons. Thus, we can expect the magnon-photon coupling strengths in the canted phase of the antiferromagnet to be
\begin{equation}
    \begin{split}
        g_{c\alpha} (H) &= K_{\alpha} (m(H) + p(H)) \sin{\theta(H)} \\
        g_{c\beta} (H) &= K_{\beta} V_{\alpha \beta} (n(H) - q(H)) \\
    \end{split}
\end{equation}
where $V_{\alpha \beta}$ explicitly takes into consideration the enhancement in coupling strength of the $\beta$ mode due to larger effective number of spins participating in it. We find this factor by calculating ratio of effective mode volumes where the field is oriented along the $z$ and the $y$ directions from finite element simulation using COMSOL Multiphysics. This gives an enhancement factor $V_{\alpha \beta} \approx \sqrt{2.67}$ since net coupling strength scales with number of spins as $\sqrt{N}$. $K_{\alpha}$ and $K_{\beta}$ are two proportionality constants that we use as fit parameters and take into account the details of the photon polarization. The coupling strength saturates on reaching the field where all the spins make transition to the collinear state. These equations clearly demonstrate field dependence of the magnon-photon coupling strength for both the antiferromagnetic modes. One can take a similar approach to calculate the coupling strengths for the antiferromagnetic state before the spin-flop transition. Although, a more detailed microscopic knowledge and control of the system is required for this, we note that the coupling strengths of the two modes in the antiferromagnetic regime can vary depending on the distribution of spins and their effective in-plane orientation at these fields with respect to the applied field direction and the easy axis (details shown in the Supplemental Material \cite{SupplMat}). We take the zero-field values of the coupling strengths as fit parameters which undergo smooth transition to canted phase dynamics beyond spin-flop transition. Using this model, we fit the avoided crossings of the antiferromagnetic resonance modes with the lower harmonic of the cavity mode (see inset of Fig.~\ref{fig:fig3} (c)). The mode dispersions of the antiferromagnetic modes used in the fitting process and calculated using the relations mentioned above are plotted in Fig.~\ref{fig:fig3} (a). The coupling strengths extracted from the fit are plotted in Fig.~\ref{fig:fig3} (b). The reflection coefficient of our hybrid magnon-photon system derived using input-output theory \cite{collettSqueezingIntracavityTravelingwave1984, schusterHighCooperativityCouplingElectronSpin2010} is given by
\begin{equation}
    \begin{split}
        S_{11} &= 1 - \frac{\kappa_{e}}{i (\omega - \omega_{c}) + \frac{\kappa_{i} + \kappa_{e}}{2} + \sum\limits_{\eta = \alpha, \beta} \frac{\abs{g_{c \eta}}^{2}}{i (\omega - \omega_{s, \eta}) + \frac{\gamma_{s, \eta}}{2}}} - \frac{2 \kappa_{e}}{i (\omega - 2 \omega_{c}) + \frac{2 \kappa_{i} + 2 \kappa_{e}}{2} + \sum\limits_{\eta = \alpha, \beta} \frac{\abs{g_{c \eta}}^{2}}{i (\omega - \omega_{s, \eta}) + \frac{\gamma_{s, \eta}}{2}}}
    \end{split}
\end{equation}
where $\kappa_{i}$ and $\kappa_{e}$ are the internal and external decay rates of the fundamental cavity mode, $\omega_{c}$ is the cavity frequency, $g_{c \eta}$ is the magnon-photon coupling strength between the cavity photons and the magnons of the antiferromagnetic mode $\eta$ (where $\eta = \alpha, \beta$), and $\omega_{s, \eta}$ and $\gamma_{s, \eta}$ are the frequencies and the decay rates of the antiferromagnetic modes respectively. Inserting the field dependent mode dispersions and coupling strengths shown in Fig.~\ref{fig:fig3}~(a) and (b) respectively in this expression we evaluate and plot the reflection coefficient as a function of frequency and magnetic field in Fig.~\ref{fig:fig3} (c). We note that this agrees with and brings out some interesting aspects of the measured response. Firstly, we note that our model explains the downward bend in lower cavity harmonic and upward bend in upper cavity harmonic near the spin-flop field. This is caused by an enhanced coupling strength between the optical magnons and cavity photons at lower fields compared to that between the acoustic magnons and photons (additional details are provided in the Supplemental Material \cite{SupplMat}). As field increases, the coupling strength between the photons and the acoustic magnons increases, whereas that between photons and optical magnons goes down until it vanishes around the field marking transition from the canted to the collinear state, when all spins transition to the collinear phase resulting in a ferromagnetic resonance like dynamics. This in turn correctly fits to the smaller avoided crossing with the optical mode as well as the downward pull to the lower cavity harmonic near this avoided crossing. In this regard, we note that the spins forming the acoustic and the optical antiferromagnetic resonance modes belong to different geometric locations due to the complex field distribution in a coplanar waveguide resonator. As a result, we expect an additional relative enhancement factor in the acoustic mode coupling strength compared to the optical mode with increasing field as all the spins ultimately orient in a collinear fashion on reaching the canted to collinear transition field. This could explain the larger avoided crossing gap in the experimental data for the higher harmonic of the cavity. We demonstrate this aspect in the Supplemental Material \cite{SupplMat}, and could be used along with a better microscopic knowledge about the system to fit to the full spectrum of the device with limited number of fitting parameters. The oscillation in the background of the signal plotted in Fig.~\ref{fig:fig2}~(b) is due to standing waves in the cables, and can be minimized using components with larger bandwidth.

\section{Weakly interacting spin waves}

\begin{figure}[H]
    \centering
    \includegraphics{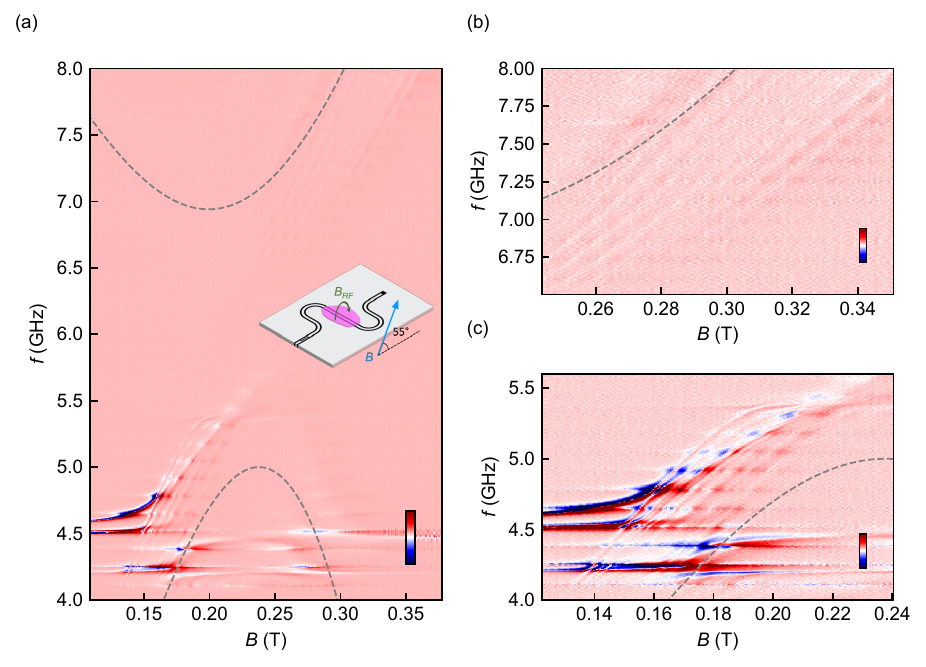}
    \caption{(a) Variation of the field derivative of the reflection coefficient $\frac{d}{dB} \abs{S_{11}}^{2} (f)$ with magnetic field for the CrCl$_{3}$ on the CPW resonator when the magnetic field is applied at an angle of 55$\degree$ with the device plane (colorbar ranges from -85 to +65 dB/T). (b) and (c) Zoomed plots showing the field dispersion near the positions of the upper and the lower hybrid AFMR modes formed due to hybridization of the optical and the acoustic magnons (colorbars range from -30 to +25 dB/T). These plots show the multiplicity of SSW modes which does not couple with the cavity mode. The gray dashed lines correspond to the field dispersions of the two hybrid antiferromagnetic modes for a sample with a CrCl$_{3}$ placed on a CPW transmission line and magnetic field applied at a similar 55$\degree$ angle.}
    \label{fig:fig4}
\end{figure}

When we apply the magnetic field at an angle of 55$\degree$ with respect to the plane of the device, we observe a modified mode dispersion due to magnon-magnon coupling between the acoustic and the optical AFMR modes \cite{macneillGigahertzFrequencyAntiferromagnetic2019, kapoorObservationStandingSpin2021}. These modes hybridize to give rise to two hybrid AFMR modes which interact with the cavity mode to give the two avoided crossings, shown in Fig.~\ref{fig:fig4}~(a). Interestingly, we observe multiplicity of standing spin wave modes formed across the crystal thickness in the vicinity of these hybrid AFMR modes which do not undergo hybridization with the cavity mode. We show this in zoomed plots in Fig.~\ref{fig:fig4}~(b) and (c) for the regions close to the upper and the lower hybrid magnon modes. The gray dashed lines show the bare hybrid magnon modes without any magnon-photon coupling, extracted from transmission spectroscopy of a CrCl$_{3}$ crystal placed on a CPW transmission line. Comparing the slopes we find the SSW modes have a slope close to the hybrid AFMR modes without cavity coupling (additional details and analysis on spin waves are provided in the Supplemental Material \cite{SupplMat}). Hence, although the AFMR modes interact resonantly with the cavity mode, the SSW modes do not. A possible reason for this is that for the SSW modes with non-zero wave vector, the alternating opposite phases across the crystal thickness reduce the effective interaction mode-volume between the SSW and the cavity modes. A more detailed micromagnetic exploration and simulation is needed to understand this aspect better. Additional field-angle dependent high frequency measurements can also provide useful insights. In magnon-based hybrid device architectures, this type of non-interacting spin waves could provide low-loss signal propagation channels in an otherwise coupled system.

\section{Conclusion}

We explored collective spin dynamics in a cavity magnonics system comprised of a CrCl$_{3}$ crystal placed at the current antinode of a CPW resonator made of NbN. We observed signature of spin-flop transition on the fundamental and the first harmonic of the cavity mode due to a stronger dispersive coupling between the optical AFMR mode and the cavity mode at low magnetic fields. Using a magnon-photon coupling model we fitted the avoided crossings and evaluated field dependence of magnon-photon coupling strength for both AFMR modes. We found a magnetic field dependent magnon-photon coupling strength which is distinctive and non-monotonic for the acoustic and the optical AFMR modes resulting from the magnetic structure change in the crystal till the point of transition to the collinear state. This provides a broader understanding of collective spin dynamics and the effect of magnetic structural change on the magnon-photon coupling strength in a van der Waals type layered antiferromagnet, along with a sensitive cavity-based method to detect magnetic phase transitions like spin-flop in an antiferromagnet. Our finding of non-interacting spin waves, alongside AFMR modes which interact resonantly with the cavity mode, provides a way to use these as low-loss propagation channels in an otherwise coupled cavity-magnonics system.

\textit{Acknowledgements} - We thank Rajamani Vijayaraghavan, Vibhor Singh, Akashdeep Kamra, Anjan Barman, Meghan Patankar, Suman Kundu, Sumeru Hazra, Sudhir Sahu, Ameya Riswadkar, Anirban Bhattacharjee, and Srijita Das for helpful discussions and experimental assistance. We acknowledge the Swarnajayanti Fellowship of the Department of Science and Technology (for M.M.D.), DST Nanomission Grant Nos. SR/NM/NS-45/2016, SERB SUPRA SPR/2019/001247, ONRG Grant No. N62909–18-1–2058, and the Department of Atomic Energy of the Government of India Grant No. 12-R\&D-TFR5.10–0100 for support.

\textit{Data availability} - The data that support the findings of this article are openly available \cite{mandal_2025_15321722}.

\onecolumngrid

\section*{Supplemental material}

\renewcommand{\thesection}{S\arabic{section}}
\setcounter{section}{0}

\renewcommand{\thefigure}{S\arabic{figure}}
\setcounter{figure}{0}

\renewcommand{\theequation}{S\arabic{equation}}
\setcounter{equation}{0}

\renewcommand{\thetable}{S\arabic{table}}
\setcounter{table}{0}

\section{Response without field dependence of coupling strength}

\begin{figure}
    \centering
    \includegraphics{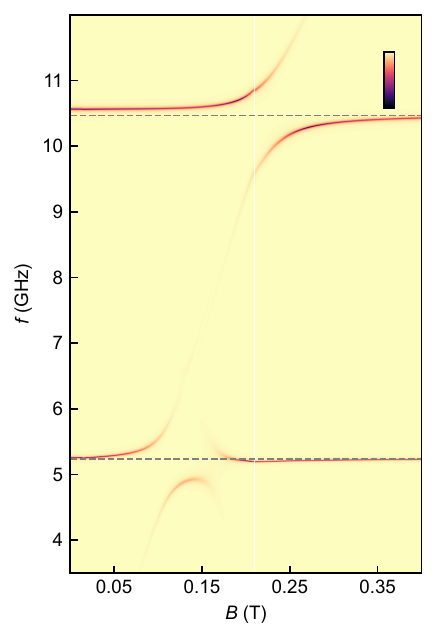}
    \caption{\label{fig:fig S1} Reflection coefficient, $|S_{11}|^{2}$, expressed in dB, of the full cavity antiferromagnet system calculated from the input-output theory using only the field dependent frequency dispersion as shown in Fig.~3~(a) of the main text and with coupling strengths $g_{\alpha}=0.37$~GHz and $g_{\beta}=0.57$~GHz (the colorbars range from -25 to 0 dB).}
\end{figure}

For the measurements described in the manuscript, we have used a single-port half-wave coplanar waveguide resonator coupled to the measurement port through a coupling capacitor for reflection measurement. Detailed characterization of this resonator, made of NbN, with the magnetic field shows its capability to retain a high quality factor and minimal frequency dispersion with magnetic field \cite{mandalCoplanarCavityStrong2020}. The CrCl$_{3}$ crystal is placed on the current antinode of the resonator to couple with the electromagnetic mode of the cavity. Typically, a constant coupling strength characterizing the avoided crossing is used for the magnetic resonance modes coupling with the cavity mode. In Fig.~\ref{fig:fig S1} we show the reflection coefficient calculated with coupling strengths extracted from the avoided crossings with a simple model that considers the magnon-photon coupling strengths to be constants \cite{mandalCoplanarCavityStrong2020}. This demonstrates the need for including the field dependence of the magnon-photon coupling strengths in the system Hamiltonian for a better understanding of the full mode dispersion of the hybrid device, and features such as the opposite-ward bends in the two cavity harmonics close to spin-flop field, a smaller avoided crossing between the fundamental cavity mode and the optical mode, and a large avoided crossing between the acoustic mode and the first harmonic of the cavity close to the field marking the transition from canted to collinear state.

\section{Hamiltonian in AF regime}

In the antiferromagnetic orientation before the spin-flop transition, the Hamiltonian can be similarly written as

\begin{equation}
    H = -\gamma \hbar \sum_{i,j} \vb{H}_0 \vdot \vb{S}_{i,j} + \sum_{i,j} 2 J_{ij} \vb{S}_{i} \vdot \vb{S}_{j} +\sum_{i,j} D_{y} (S_{i,j}^{y})^{2} - \sum_{i,j} D_{z} (S_{i,j}^{z'})^{2}
\end{equation}

where the spin operators can be expressed in terms of spin wave excitation or magnon operators as $S_{i}^{x'} = \sqrt{\frac{S}{2}} (a_{i} + a_{i}^{\dag})$, $S_{i}^{y'} = \frac{1}{i} \sqrt{\frac{S}{2}} (a_{i} - a_{i}^{\dag})$, $S_{i}^{z'} = S - a_{i}^{\dag} a_{i}$, and $S_{j}^{x''} = \sqrt{\frac{S}{2}} (b_{j} + b_{j}^{\dag})$, $S_{b}^{y''} = - \frac{1}{i} \sqrt{\frac{S}{2}} (b_{j} - b_{j}^{\dag})$, $S_{j}^{z''} = S - b_{j}^{\dag} b_{j}$. Using a procedure similar to that described in the main text, this Hamiltonian can be simplified to

\begin{equation}
    H = (A_{a} a^{\dag} a + A_{b} b^{\dag} b) + B (a b + a^{\dag} b^{\dag}) + \frac{C}{2} (a^{2} + b^{2} + \textrm{h.c.})
    \label{eq:HAF}
\end{equation}

where

\begin{equation}
\begin{split}
    A_{a} &= A + \gamma H_{0} \cos{\phi} \\
    A_{b} &= A - \gamma H_{0} \cos{\phi} \\
    A &= \gamma \left[H_{E} + \frac{H_{Ay}}{2} + \frac{H_{Az}}{2} \left( 2 \cos^{2}{\phi} - \sin^{2}{\phi} \right) \right] \\
    B &= \gamma H_{E} \\
    C &= -\gamma \left( \frac{H_{Ay}}{2} + \frac{H_{Az}}{2} \sin^{2}{\phi} \right)
\end{split}
\end{equation}

where $\phi$ is the azimuthal angle in the plane of the sample made by either of the sublattice moment with the direction of the applied field. The eigenfrequencies of the two modes in the antiferromagnetic regime are given by $\omega_{\alpha, \beta} = \sqrt{A^{2} + (\gamma H)^{2} - (B^{2} + C^{2})} \pm 2 \sqrt{(A^{2} - B^{2}) (\gamma H)^{2} + B^{2} C^{2}}$. We note that Eq.~\ref{eq:HAF} takes a form akin to Eq.~3 in the main text, which can be diagonalized easily, only for $H \rightarrow 0$. In that case, it can be shown that $m_{y} \sim (m - p) (\alpha - \alpha^{\dagger})$, $m_{x} \sim m_{z} \sim (n + q) (\beta + \beta^{\dagger})$ where $m = \sqrt{\frac{A + \omega_\alpha}{4 \omega_\alpha}}$, $n = \sqrt{\frac{A + \omega_\beta}{4 \omega_\beta}}$, $p = \sqrt{\frac{A - \omega_\alpha}{4 \omega_\alpha}}$ and $q = \sqrt{\frac{A - \omega_\beta}{4 \omega_\beta}}$ with $\omega_{\alpha, \beta} = A^{2} - (B \mp C)^{2}$ which are dependent on $\phi$. Since in-plane behavior of the antiferromagnet depends on local relative orientations of the domains and the competition between the in-plane anisotropy and applied field directions, a more controlled experiment and knowledge about the microscopic details about the crystal magnetization is required to precisely know the in-plane orientation distribution of the moments and hence the exact coupling strengths.

\section{Enhanced acoustic magnon-photon coupling and bend in second harmonic of cavity}

\begin{figure}
    \centering
    \includegraphics{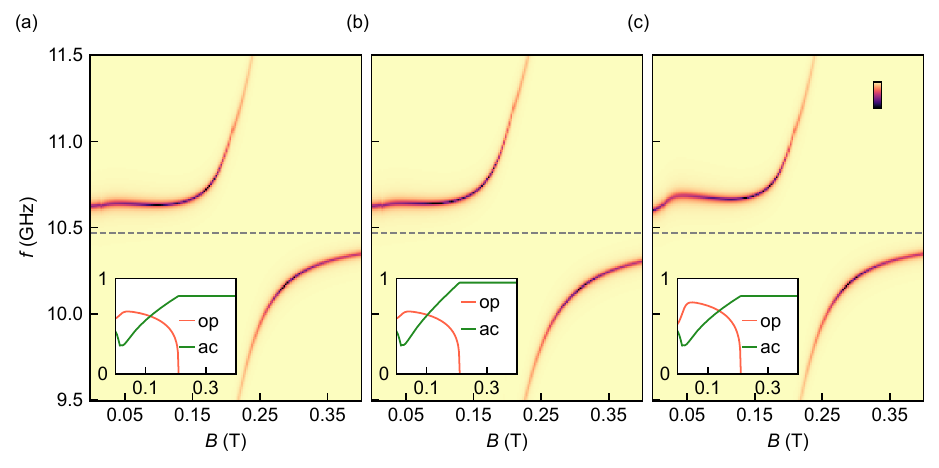}
    \caption{(a), (b) and (c) Reflection coefficient $|S_{11}|^{2}$ of the cavity-antiferromagnet system, expressed in dB, calculated from the input-output theory using field-dependent frequency and coupling strengths of the acoustic and the optical AFMR modes. (a) Calculation based on the proposed model. (b) Calculation with an additional enhancement in the acoustic magnon-photon coupling strength due to an increase in the number of participating spins closer to ferromagnetic transition, which shows an increase in the avoided crossing gap. (c) Calculation with an additional enhancement factor for the magnon-photon coupling strength of the optical mode over the acoustic mode close to the spin-flop transition, which shows a larger bend of the second harmonic of the cavity mode close to the point of transition into the canted orientation (the colorbar range from -25 to 0 dB).}
    \label{fig:fig S2}
\end{figure}

In this section, we discuss two subtle features visible in the experimental data and their possible explanation. Firstly, we notice that the avoided crossing gap for the second cavity harmonic as given by our model is smaller than that observed experimentally. One possible way to explain that is by noticing that as the moments start aligning in collinear orientation closer to the transition field marking canted to collinear transition, the number of spins that were taking part in the formation of the acoustic mode should also increase because the acoustic AFMR mode ultimately transforms to the FMR mode. As a result, the net coupling strength should have an additional increase factor which should be proportional to $\sqrt{N}$ with $N$ being the number of spins. We assume a linear increase in the number of participating spins in the formation of the acoustic mode from all the regions that were contributing exclusively to the formation of the optical mode close to the spin-flop field. The mode dispersion close to the second cavity harmonic with and without taking into account this additional factor is shown in Fig.~\ref{fig:fig S2}~(a) and (b) respectively, with the insets showing the corresponding magnon-photon coupling strengths from the model. We observe that inclusion of this factor indeed increases the avoided crossing gap and approaches the experimentally observed value. Secondly, the initial upward bend in the second harmonic of the cavity near the spin-flop transition is higher in the experimental data. A possible explanation for this is that before and close to the spin-flop transition, the exact magnon-photon coupling strengths of the optical and acoustic AFMR modes with the cavity depend on the in-plane orientation and distribution of the magnetic moments. By including a phenomenological parameter denoting an initial relative enhancement in coupling strength for the optical mode compared to the acoustic mode over the values estimated from the fit, we find that indeed a larger bend in the second harmonic is possible. This has been demonstrated in Fig.~\ref{fig:fig S2}~(c) along with the field dependence of the magnon-photon coupling strengths in the inset. However, a more detailed micromagnetic knowledge about the initial spin orientation in the crystal plane is required for a quantitative understanding of this enhancement factor. This can also enable further refinement of the proposed model to do a full self-consistent fit to the entire mode dispersion.

\section{Angle dependence of the hybrid modes}

\begin{figure}
    \centering
    \includegraphics{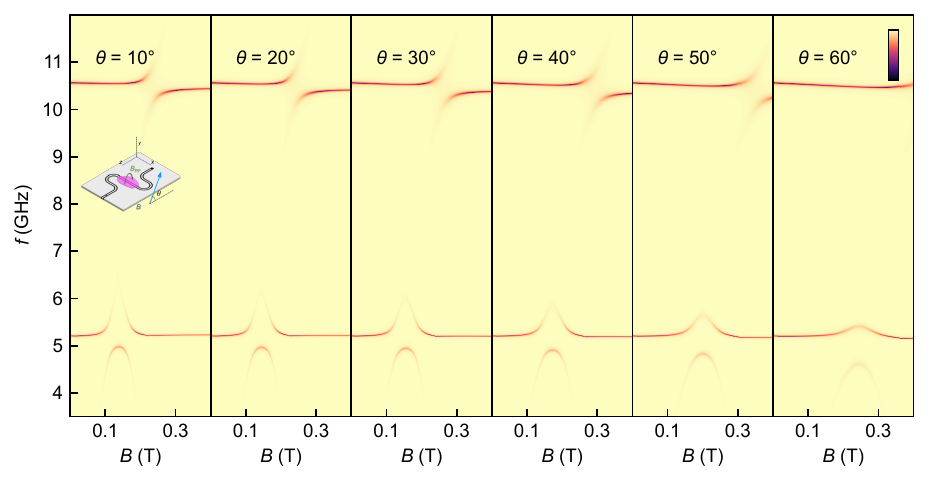}
    \caption{Variation of the reflection coefficient $|S_{11}|^{2}$ of the cavity-antiferromagnet system calculated from the input-output theory with angle of the external magnetic field (the colorbar range from -25 to 0 dB). The inset in first figure shows the direction of the external magnetic field with respect to the sample plane.}
    \label{fig:fig S3}
\end{figure}

The angular dependence of the cavity-antiferromagnetic hybrid modes has been shown in Fig.~\ref{fig:fig S3}. As the external magnetic field makes a non-zero angle with the sample plane, the vertical component introduces a magnon-magnon coupling. This hybridizes the acoustic and the optical AFMR modes, resulting in an upper and a lower hybrid magnon mode. These modes undergo hybridization with the cavity mode, resulting in hybrid cavity-antiferromagnetic modes. A proper estimation of the coupling strengths of the hybrid antiferromagnetic modes with the cavity mode depends on the interaction of the transforming spins with the cavity photons under the effect of an additional magnon-magnon coupling. Nevertheless, a similar calculation based on the input-output theory explained in the main text with an average coupling strength for the modes, shows that the mode dispersion expected from the theory is in good agreement with the experimental data for a non-zero angle. A detailed micromagnetic understanding of the system will help in precisely capturing the variation of the coupling strength with field for both the hybrid modes as a result of magnon-magnon coupling.

\begin{figure}
    \centering
    \includegraphics{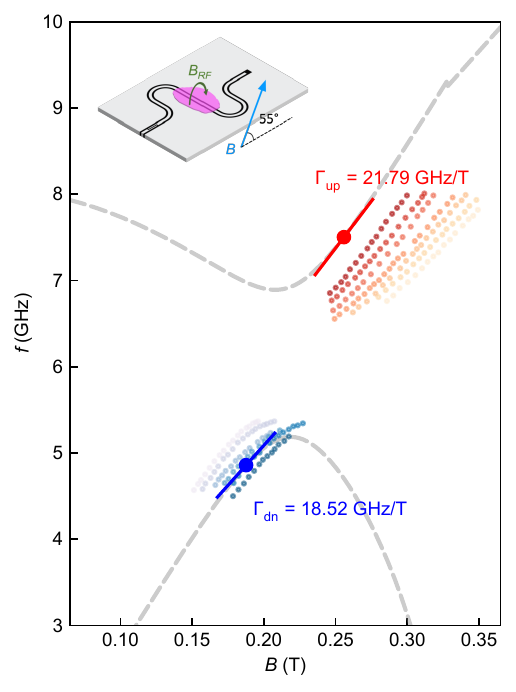}
    \caption{Comparison of slopes of the standing spin wave modes with the associated AFMR modes for an external field angle of 55$\degree$ with the sample plane. The gray dashed lines show the AFMR mode dispersion calculated using the two-sublattice model. The red and the blue multiplicity of modes shown are the standing spin wave modes found from the experiment, associated with the upper and the lower hybrid AFMR modes, respectively.}
    \label{fig:fig S4}
\end{figure}

\section{Nature of standing spin waves}

The standing spin wave (SSW) modes do not undergo substantial hybridization with the cavity mode, and follow a dispersion close to the hybrid AFMR modes as a result of magnon-magnon coupling between the acoustic and the optical AFMR modes. This is clear in the Fig.~\ref{fig:fig S4} where the dispersion of the hybrid AFMR modes from the two-sublattice antiferromagnetic model for CrCl$_{3}$ has been plotted along with experimentally obtained SSW dispersions with magnetic field. Although a higher signal-to-noise ratio is required for proper indexing of the SSW modes, we find that the average spacings between them are 7.9~mT for the upper branch and 5.5~mT for the lower branch, around the chosen AFMR mode frequencies 7.50~GHz and 4.86~GHz for the upper and the lower AFMR modes, respectively. From a comparative analysis of the spin wave models, the spacing $\Delta H$ between the SSW modes depends on the effective distance between the SSW mode boundaries ($L_\textrm{eff}$) and the slope of the associated magnetic resonance mode ($\Gamma$) as $\Delta H \propto (L_\textrm{eff}^{2} \Gamma)^{-1}$ according to Kittel model, and $\Delta H \propto (L_\textrm{eff}^{2} \Gamma)^{-1/2}$ according to Portis model \cite{kapoorObservationStandingSpin2021, kittelExcitationSpinWaves1958, portisLOWLYINGSPINWAVE1963}. The extracted $\Gamma_\textrm{up}/\Gamma_\textrm{dn}$ from Fig.~\ref{fig:fig S4} is 1.17. Now, comparing the corresponding ratios of the spacings between the spin waves, we get $\Delta H_{dn}/\Delta H_{up}=$ 0.69 and $(\Delta H_{dn}/\Delta H_{up})^{2}=$ 0.47. This suggests that the effective length between the boundaries of the SSW modes associated with the upper AFMR branch is lower compared to the ones associated with the lower branch. This demonstrates a formalism to study the efficiency of propagation of the spin waves with frequency and magnetic field.

\end{document}